\documentclass[11pt]{amsart}
\usepackage{geometry}                % See geometry.png to learn the layout options. There are lots.
\usepackage{graphicx}
\usepackage{amssymb}
\usepackage{epstopdf}

\graphicspath{
	{plots/}
}

\usepackage[colorlinks = true,
            linkcolor = blue,
            urlcolor  = blue,
            citecolor = blue,
            anchorcolor = blue]{hyperref}

\usepackage[ruled,vlined]{algorithm2e}

\hypersetup{
    colorlinks=true,
    linkcolor=blue,
    filecolor=magenta,      
    urlcolor=cyan,
}
 
\title{Latent Channel Networks}
\author{Clifford Anderson-Bergman }
\author{Phan Nguyen} 
\author{Jose Cadena Pico}

\newcommand{\edge}{e}
\newcommand{\latentedge}{\tilde e}
\newcommand{\latcon}[3]{c_{#1#2#3} }
\newcommand{\hubProb}{\theta}

\newcommand{\expCon}{C}
\newcommand{\plotSize}{14cm}

\usepackage{xcolor}

\begin{document}

\maketitle

\section{Abstract}

% Latent Euclidean embedding models a network by representing each node as a point in a Euclidean 
% space wherein the probability of an edge between two nodes is inversely related to 
% the distance between the corresponding points in the embedding. 
% This implies that for two nodes to share an edge with high probability, they must be relatively close in 
% \emph{all} dimensions. This constraint may be overly restrictive for describing modern networks in which having strong similarities 
% in \emph{at least} one area may be sufficient for a high edge probability. 

 In modern social networks, individuals often belong to several communities
 with widely varying number of connections made through each community. 
 To capture this social structure, we introduce the 
 Latent Channel Networks model, 
 in which nodes share observed edges if they make a connection through at least one latent channel. 
% This results in a network model that shares a similar structure to a competing model,
% so we compare and contrast the two models. 
We present the mathematical model, interpretations of various parameters
 and an EM algorithm whose computational complexity 
 is linear in the number of edges and channels and 
 can handle a graph in which some of the edge statuses may be unknown. 
 We compare the performance of our model to a similarly structured competing model on both
  synthetic and and a subsample of the Facebook100 graphs. 
 Although both models show strong predictive capabilities, 
 we find that our model tends to suffer less from overfitting and consistently achieves 
 a moderate improvement in out-of-sample AUC when estimating masked edge status on the Facebook data. 
 We discuss insights the model provides into the graph structure of the Facebook data. 

\section{Introduction}
% \subsection{Definitions}
% \NOTE{Cut this subsection? And move assumptions to the beginning of section 3}
% In this work, we define $G = (N, E)$ to be a graph with a set of nodes $N$ and 
% an adjacency matrix $E$ such that $E_{ij} = 1$ if nodes $i$ and $j$ share an 
% edge and $E_{ij} = 0$ otherwise. We focus mainly undirected graphs, implying $E_{ij} = E_{ji}$
% and ignore loops, implying $E_{ii} = 0$.
% The degree of a node is defined as the number of edges attached to it. 
% A classic example of this is a social network in which nodes represent 
% individuals and edges exist between two individuals if they are listed as friends.
% Another common example is a co-authorship graph in which nodes represent researchers 
% that share an edge if they have co-authored a paper.

\subsection{Relevant Work}
In the analysis of graph data, a common goal is describing a network in a reduced-order 
space that provides insight into the underlying graph structure. 
One of the simplest structures is the stochastic block model (SBM) \cite{sbm}. 
In this model, each node belongs to an unobserved block, and nodes
have a fixed probability of having an edge with nodes within their block
and another fixed probability of having an edge with nodes outside their block. 
Typically, the within-block edge probabilities are greater than the between-block edge probabilities, so nodes are more likely to 
share an edge with nodes within the same block, and each block may be considered a cluster or community. 
Recent work covers efficient estimation of the parameters of SBMs
\cite{sbmExact,sbmSpec}, time-evolving or dynamic SBMs \cite{dsbm},
 statistical characteristics of the estimators \cite{specConsist,sbmAsym}
 model selection \cite{sbmSelection} and hierarchical SBMs \cite{nestedsbm}.

One disadvantage of SBMs is that the expected within-block degree of a node is constant, with a variance implied by a binomial distribution. 
This fails to capture a commonly observed phenomenon in social networks; often, a small number of nodes express an extremely high degree relative to most other nodes. 
Other models have been proposed to capture this property, such as the degree-corrected SBM \cite{dcsbm}, in which edge probabilities are based on block membership
 \emph{and} a given node's degree. 

Another limitation of SBMs is that it is \emph{hard clustering} approach: each node deterministically belongs to a single block.
Several alternatives have been considered that allow for \emph{soft clustering}, including the mixture SBM \cite{msbm} in which each node belongs to a each block 
with a given probability. Another approach is to 
maximize the modularity score \cite{modularity}, but with community membership 
described as a probability vector rather than as a categorical variable \cite{softmod1,softmod2,softmod3}.
Other metrics such as the overlapping correlation coefficient \cite{occ} may also be used. 

Ball, Karrer, and Newman presented a soft clustering-related model that uses overlapping clusters \cite{bkn}; 
we refer to this model as the BKN model.
In this model, several latent communities exist, 
and each node has a community intensity parameter associated with each community. 
The number of edges between two nodes is modeled as a Poisson distribution
with a mean given by the sum over all communities of the product of community intensities 
between the two nodes. Although this modeling assumption implies counts rather 
than binary values, the authors present the Poisson distribution as a reasonable approximation 
when edges are binary.

The Euclidean embedding model \cite{eucEmbed} is another popular probabilistic model.
In this model, each node is represented as a point in a latent Euclidean space, 
with edge probabilities being inversely proportional to distances.
Because these probabilities are directly modeled, one can naturally 
allow them to be a functions of both latent distance
\emph{and} linear predictors associated with each node. 
This model also allows for both very high and low degree
nodes; each node has its own intercept, and high degree nodes are simply 
nodes whose intercept is exceptionally high. 
Traditional MCMC approaches were initially presented for inference, 
and methods to accelerate this have also been proposed, including using variational Bayes \cite{vbEuc}
and stratified case-control sampling \cite{stratEuc}.
A similar model is the random dot product graph \cite{randDot1,randDot2},
in which nodes are represented in a latent space and edge probabilities
between two nodes are given by the dot product of their latent positions. 
These positions can be estimated via eigendecompositions of
the adjacency matrix \cite{eigenDot}. While clusters are not explicitly 
modeled in latent space embedding, clustering may still be performed 
on the lower-dimensional latent embedding. 

One last class of models that is related to our work is the 
sender/receiver model \cite{sendReceive1,sendReceive2}. 
These models are applied to directed graphs, and each node has a 
parameter that controls how frequently it broadcasts and receives connections. The probability 
of an edge from a source node to target node is then a function 
of the source's broadcast strength and the target's receiving strength. 

\subsection{Latent Channel Network}
One major disadvantage of an Euclidean embedding is that 
two nodes must be close in \emph{all}
dimensions in order to have a high edge probability. The suitability of this assumption has been called into question 
for web-based applications \cite{euclidBadInternet} and legislative voting patterns \cite{sphericalEmbed}.
In modern social networks, being similar in \emph{at least one} 
social dimension may be sufficient for high edge probability. 
For instance, \cite{tweets} demonstrated that 
while political retweeting falls very tightly along political lines, 
it is \emph{not} strongly predictive of retweeting of non-political topics.
While one could relax this requirement by embedding 
the network in a non-Euclidean space, say $L^p$, a few issues arise; 
picking $p$, unstable and computationally expensive likelihoods and interpretation of the parameters. 
This shortcoming motivated us to work on a new model that 
more naturally allows two nodes to share a connection if they share
\emph{some} social overlap.

To capture this, we present the Latent Channel Network (LCN) model.
Here, we assume that two nodes will share an edge in the graph if they are connected through 
\emph{at least one} of potentially several unobserved latent channels.
The probability of two nodes connecting through a given channel 
is the product of each node's frequency of use of the given channel. 
Thus, two nodes may differ strongly in many social areas, \emph{i.e.} 
they do not need to frequently use \emph{all} of the same channels, 
yet the model still allows them to have a high edge probability if they both frequently communicate 
through one or more shared channels.

\subsection{Relation to Previous Work}

The LCN  is closely related to the BKN model \cite{bkn}. 
The fundamental difference between the two models is that
rather than modeling edge counts as a Poisson variable with mean equal to the sum of means across
all channels, the LCN instead models binary edges 
as a Bernoulli distribution, where the probability of having an edge is 
one minus the probability of having no edges through any of the channels.
In Section \ref{sec:BKN}, we show that this produces a model that encourages overlap in channels between edge-pairs, 
while the BKN model discourages edge-pairs in the model from having strong connections through
multiple channels, and we empirically show that our model leads to better edge predictions on social network data. 

If one considers channels to represent communities, both our model and the BKN model can be viewed 
as similar to an overlapping communities model with one key difference. 
In particular, in the formal definition overlapping communities provided in \cite{overlapReview}, 
each node has a probability vector 
containing probabilities of association to each community which sums to one for each node.
In contrast, both the LCN and BKN
community membership parameters are attachment parameters for independent communities, 
and thus the vector of probabilities/intensities is not constrained to sum to one for each node. 
Thus, being strongly attached to one community does not prevent 
a node from being strongly attached to another community. 
These models naturally allow networks
that contain a mix of high degree nodes (those that use multiple channels with high frequency)
and low degree nodes (those that use all channels with low frequency). 
%This differentiates from standard soft clustering, 
%as being strongly connected to one community does \emph{not} constrain
%a node from being strongly connected to another community. 

One can also view the LCN model as a modified sender/receiver model, 
in which each node has multiple opportunities to make a connection and direction is ignored. 

\subsection{Structure of Paper}

In Section \ref{sec:model}, we formally define our model  and present various ways
to interpret meaningful parameters from the model. We also compare and contrast our model with the BKN model. 
In Section \ref{sec:algorithm}, we present 
both a simple and more computationally efficient algorithm to compute the maximum 
likelihood estimate of the model parameters. In Section \ref{sec:applications}, we apply the model 
to synthetic data and several graphs from the Facebook100 dataset \cite{fb100}. In Section \ref{sec:discuss}, 
we review our work and discuss potential further directions. 

\section{Latent Channel Model}
\label{sec:model}
\subsection{Model Parameterization}

\begin{figure}
\centering
\begin{minipage}{0.5\textwidth}
\includegraphics[width = 6cm, height = 8cm]{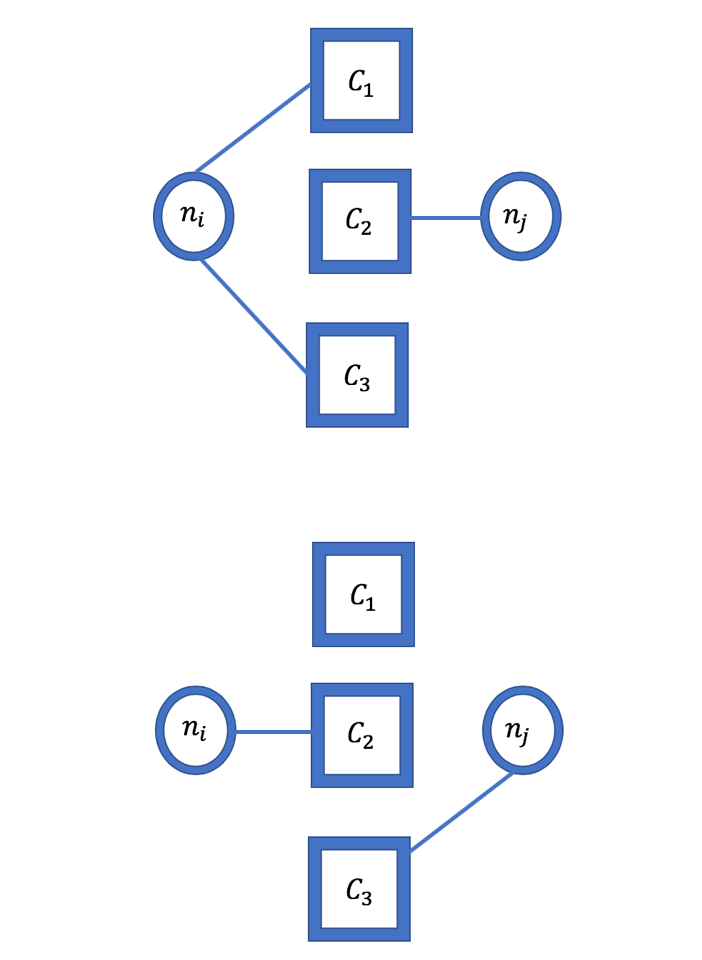}
\caption{Nodes $n_i$ and $n_j$ do not share an edge as they are not connected through any channel.}
\label{fig:unconnected}
\end{minipage}%
\begin{minipage}{0.5\textwidth}
\includegraphics[width = 5cm, height = 8cm]{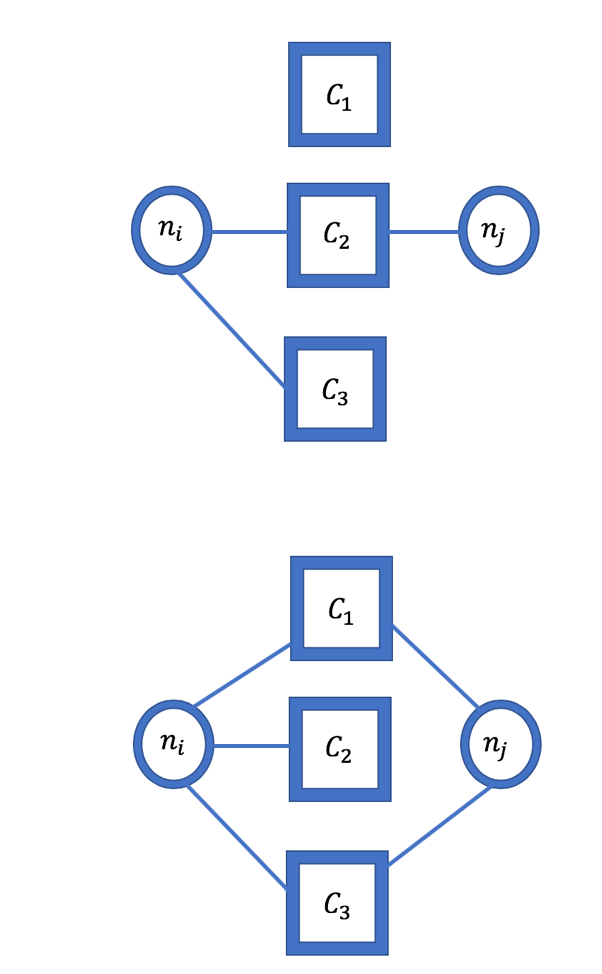}
\caption{Nodes $n_i$ and $n_j$ share an edge as they are  connected through at least one channel.}
\label{fig:connected}
\end{minipage}
\end{figure}

Let $G$ be undirected graph with nodes $n_1$,...,$n_{N_n}$ and edges $\edge_{ij} = 1$ if $n_i$ and $n_j$ are connected and 0 otherwise. 
For simplicity, we assume the graph has no self-loops.
Let $N_n$ to be the number of nodes and $N_e$ to be the number of edges of the graph.
We augment this observed graph with a set of latent channels $C_1$,\ldots,$C_K$ that provide intermediate connections between nodes. 
In particular, we introduce latent edges $\latentedge_{ikj}$, which is equal to 1 if node $n_i$ shares a latent edge to channel $C_k$ toward node $n_j$ and 0 otherwise.
Our model then dictates that a pair of nodes share an observed edge if they are both fully connected through one or more latent channels. More formally, 
\[
\edge_{ij} = 
\begin{cases}
1 & \text{if there exists } k \text{ such that } \latentedge_{ikj} = \latentedge_{jki} = 1 \\
0 & \text{otherwise}.\\
\end{cases}
\]
Examples of this architecture are illustrated in Figures \ref{fig:unconnected} and \ref{fig:connected}. For simplicity, we define
\begin{equation}\label{eq:hubConDef}
\latcon{i}{j}{k} = \mathbb{I}( \latentedge_{ikj} = \latentedge_{jki} = 1 ).
\end{equation}
In other words, $\latcon{i}{j}{k}$ is an indicator that nodes $n_i$ and $n_j$ are connected through channel $C_k$. 
In general, $\latentedge_{ikj}$ and $\latcon{i}{j}{k}$ cannot be observed directly. 
However, our model assumes that for all $n_j$, $\latentedge_{ikj}$ are independently distributed Bernoulli 
distributions with probability $p_{ik}$. 
Thus, the marginal probability
that $n_i$ will share a edge with $n_j$ through channel $C_k$ is $p_{ik} p_{jk}$. 
To determine the probability that nodes $n_i$ and $n_j$ share an edge, we compute
\begin{equation} \label{eq:edgeProb}
\begin{split}
P(e_{ij} = 1)    & = 1 - P(e_{ij} = 0) \\
			& = 1 - \prod_{k = 1}^K \left(1 - P( \latcon{i}{j}{k}) \right) \\
                         & = 1 - \prod_{k = 1}^K (1 - p_{ik} p_{jk}). \\
\end{split}
\end{equation}
The log-likelihood of a latent channel graph can then be written as 
\begin{equation}\label{eq:llk}
L(G | p) = \sum_{i = 2}^n \sum_{j = 1}^{i-1} 
    e_{ij} \log \left( 1 - \prod_{i = 1}^K (1 - p_{ik} p_{jk}) \right) + 
    (1 - e_{ij}) \log \left( \prod_{i = 1}^K (1 - p_{ik} p_{jk}) \right).
\end{equation}
%
% We note that an undirected sender/receiver model \cite{sendReceive1} 
% is a special case of a LCN with $K = 1$.

\subsection{Interpretation of Model}

If channels correspond to latent communities, then $p_{ik}$ informally represents 
the strength of node $n_i$'s attachment to community $C_k$. However, this parameter alone can be difficult to interpret, 
as it is unclear how large $p_{ik}$ should be to be considered a strong connection. 
To aid in the interpretation of the model, we present several derived values. 

We first consider the parameter $\hubProb_{ijk}$:
\begin{equation} \label{eq:condProb}
\hubProb_{ijk} = P( \latcon{i}{j}{k} = 1 | e_{ij} = 1) = \frac{p_{ik} p_{jk} } { 1 - \sum_{k = 1}^K (1 - p_{ik} p_{jk}) }.
\end{equation}
$\hubProb_{ijk}$ represents the probability that nodes $n_i$ and $n_j$ are connected through channel $C_k$, 
given that the graph contains an edge between $n_i$ and $n_j$. This is especially relevant in the case that 
channel $C_k$ has a meaningful interpretation, such as the attachment strength parameter $p_{ik}$ being correlated with meta-data
on the nodes. For example, if the attachment strength to channel $C_k$ is strongly associated with nodes whose occupations are statisticians and 
$\hubProb_{ijk}$ is high, this suggests that given that nodes $n_i$ and $n_j$ share a connection, they have a high probability 
of having an edge through the statistical community. 
It is worth noting that 
\begin{equation}\label{eq:condProbIneq}
\sum_{k = 1}^K \hubProb_{ijk} \geq 1,
\end{equation}
and typically with strict inequality. This is because two nodes that share an edge must share \emph{at least} 
one edge through a latent channel, but may share many. For example, if channel $C_k$ represents the statistical community 
and $C_{k'}$ represents associations through a given research institution, 
statisticians at the same institution are likely to be connected through both channels $C_k$ and $C_{k'}$. 

Next, we consider 
\begin{equation}\label{eq:hubsize}
S_k = \sum_{i = 1}^{N_n} p_{ik},
\end{equation}
which we refer to as the \emph{size} of the channel. 
To interpret this parameter, note that if a new node $n_{i'}$ were to 
be fully connected to channel $C_k$, i.e. $p_{i' k} = 1$, it is expected to have 
$S_k$ connections through $C_k$. More generally, the expected number of connections
for a new node is $p_{i' k} S_k$. 

Another particularly useful parameter is $\expCon_{ik}$:
\begin{equation}\label{eq:expConnects}
\expCon_{ik} = \mathbb{E}\left[ \scriptstyle \sum_{j \neq i} \latcon{i}{j}{k} | G \right] = \displaystyle \sum_{j \neq i} e_{ij} \hubProb_{ijk}.
\end{equation}
$\expCon_{ik}$ represents the expected number of connections node $n_i$ has through channel $C_k$, 
conditional on the edges observed in the graph. 
While $p_{ik}$ tells us the strength of attachment node $n_i$ has to 
channel $C_k$, it is not sufficient to determine how many connections
node $n_i$ may have through channel $C_k$. For example, a strong attachment to a small channel
may result in fewer edges than a weak attachment to a large channel.
As such, this statistic can provide insight into the number of connections a node has through a given 
community, which is a function of both that individual's strength of attachment to the community 
\emph{and} the size of the community. 

Similar to Equation \ref{eq:condProbIneq}, we note that 
\begin{equation} \label{eq:expIneq}
\mathbb{E}\left[ \scriptstyle \sum_{j \neq i} \latcon{i}{j}{k} | G \right]  \geq \sum_{j \neq i} e_{ij},
\end{equation}
or that for node $n_i$, the expected sum of connections through \emph{all} 
channels is typically greater than the sum of all observed edges in the graph associated with that node. 
Again, this is because a single edge can be the result of connections through multiple channels. 

Caution should be taken in interpreting such parameters based on fitted data. 
As is the case for many probabilistic network models, 
we currently estimate the parameters via maximum likelihood estimation.
Given the high dimensional parameter space, 
standard asymptotic normality results 
should not be considered a reliable method for estimating uncertainty.
Therefore, we suggest using these methods for exploratory data analysis
rather than making strong inference statements about a given network. 
Alternatively, Bayesian methods may be used to determine uncertainty. 
However, to do so, one must first address the unidentifiability issue
that arises due to label switching of the channels. 

\subsection{Comparisons with BKN Model}
\label{sec:BKN}

In the LCN model, each edge in the graph is independently distributed as 
\begin{equation} \label{eq:edgeDist}
e_{ij} \sim \text{Bernoulli}\left( p = 1 - \prod_{k = 1}^K (1 - p_{ik} p_{jk} ) \right),
\end{equation}
while the BKN model assumes that it is independently distributed as 
\begin{equation} \label{eq:BKNedgeDist}
e_{ij} \sim \text{Poisson}\left( \lambda =  \sum_{k = 1}^K \theta_{ik} \theta_{jk} \right).
\end{equation}

While both models have similar structures, they have important fundamental differences.
Most notably, LCNs can only be used with binary edges, whereas the BKN can handle edge counts.
Therefore, when the data contains edges counts, the BKN model should be preferred. 

When the graph contains binary edges, 
the BKN model can be seen as a useful approximation. Indeed, the authors present it as 
such:
\begin{quotation}
``However, allowing multiedges makes the model
enormously simpler to treat [when edges should be binary]
 and in practice the number of
multiedges tends to be small, so the error introduced is also
small, typically vanishing as $1/n$ in the limit of large network
size.\cite{bkn}"
\end{quotation}
In the light of this, we view the LCN as 
as a model for treating the edges exactly as Bernoulli data, rather than using a Poisson approximation.
We note that the EM algorithm we present has the same computational complexity as that of BKN's algorithm. 

In addition to philosophical motivations, the two models also differ in how they treat multiple channel usage. 
The BKN model \emph{discourages} two nodes that share an 
edge from being too strongly attached through many channels. In particular, because
the mean number of edges is $\sum \theta_{ik} \theta_{jk}$, the log-likelihood contribution 
of a single edge is maximized when $\sum  \theta_{ik}  \theta_{jk} = 1$. 
In contrast, for the LCN, the log-likelihood contribution of a single edge is 
non-decreasing with $p_{ik}p_{jk}$, although as a single $p_{ik} p_{jk}$ approaches $1$, 
the derivative of the other $p_{ik'} p_{jk'}$ approaches $0$. Therefore, 
the BKN model discourages node pairs from being too strongly attached to 
the same channels if this leads to $\sum  \theta_{ik}  \theta_{jk} > 1$ ,
while the LCN does not penalize two nodes 
that share an edge from being overly similar. 

Another difference between the two models is the BKN model allows
one node to force another node to use a given channel. If $\theta_{ik}$ is very small, 
one might assume that this implies that node $i$ infrequently uses channel $C_k$. 
However, if $\theta_{jk}$ is very large, 
then the expected number of edges between nodes $i$ and $j$ 
through channel $C_k$ may still be large as $\theta_{ik}\theta_{jk}$
 is unbounded. 
In contrast, with the LCN model, the expected number 
of edges through channel $C_k$ between nodes $i$ and $j$ is 
$p_{ik} p_{jk}$, which is bounded above by $p_{ik}$. 
 
 Finally, we note that edge probability estimates, both latent and observed, are more interpretable in our model with binary edges. 
 For example, we can directly compute the observed edge probabilities from our model if we wish 
 to make inferences about the existence of an edge between two nodes. 
 With the BKN model, it is not quite clear how to interpret the edge predictions. 
 To illustrate, suppose the BKN model estimates that the expected edge 
 count between two nodes is 1. 
 If we interpret this as the expected value of a binary distribution, this is equivalent
 to saying that there exists an edge with probability one. 
 However, if we interpret it as a parameter of a Poisson distribution, 
 then the probability of one or more edges is $1 - e^{-1}$. Similarly, it is not clear
 how to interpret the difference between estimated edge counts greater than 1 in a BKN model.   
 Finally, if one were to use a BKN model to generate a synthetic binary network, 
 one would need to truncate any edge counts greater than 1, 
 which would induce bias when estimating the parameters with a BKN model that does not account for the truncation.

\section{Algorithm}
\label{sec:algorithm}

We appeal to maximum likelihood estimation to estimate the values of $p_{ik}$. 
In general, the problem is non-identifiable and highly 
non-concave. We will use an EM algorithm \cite{EM} to fit the parameters of the model. 

\subsection{Fundamental EM Algorithm}
To determine the steps of the EM algorithm, we make several observations. First, we note that if the values of $\latentedge_{ikj}$ were known, the log-likelihood would be simplified to 
\begin{equation} \label{eq:compllk}
L(G, \latentedge | p) = \sum_{i = 1}^{N_n} \sum_{j \neq i}^{N_n} \sum_{k = 1}^{K} \latentedge_{ikj} \log( p_{ik} ) + (1 - \latentedge_{ikj}) \log(1 - p_{ik}),
\end{equation}
which is maximized when
\begin{equation}\label{eq:Mstep}
\hat p_{ik} = \sum_{j \neq i}^{N_n} \latentedge_{ikj}  / (N_n - 1),
\end{equation}
providing our M-step in the 
EM algorithm. For the E-step, we recognize that 
\begin{equation}
\displaystyle
P(\latentedge_{ikj} = 1 | e_{ij} = 1) = 
	\frac{ p_{ik} p_{jk} + p_{ik} (1 - p_{jk}) \left( 1 - \prod_{k' \neq k} ( 1 - p_{ik'} p_{jk'} ) \right)  } { 1 - \prod_{k = 1}^K( 1 - p_{ik} p_{jk}) },
\end{equation}
\begin{equation}
P( \latentedge_{ikj} = 1 | e_{ij} = 0) = 
	p_{ik} - p_{ik} p_{jk}.
\end{equation}
For clarity, we first present a simple but computationally inefficient implementation in Algorithm \ref{alg:simple}. 
Noting that computing 
$P( \latentedge_{ikj} | e_{ij} = 1)$
requires $O(K)$ operations and 
$P( \latentedge_{ikj} | e_{ij} = 0)$ 
requires $O(1)$ operations,  this implementation then requires 
$O(N_e K^2 + (N_n^2 - N_e)K)$ computations per iteration. 
\begin{algorithm}
\SetAlgoLined
\KwResult{Fixed point estimate of $N \times K$ matrix $p$}
 Adjacency Matrix $e$; K\;
 N = nrow(e)\;
 $p$ = RandomUniform(min = 0, max = 1, nrow = N, ncol = K)\;
 maxIters = 1,000; iter = 0;
 tol = $10^{-4}$; maxDiff = tol + 1\;
 \While{ iter $<$ maxIters \& tol $>$ maxDiff}{
	iter++\;
 	\For{ i in 1:N }{
		\For{ k in 1:K }{
			\For{ j in 1:N}{
	 			$\latentedge_{ijk} = 
				\begin{cases} 
				0 & \text{ if } i = j \\
				P( \latentedge_{ikj} | e[i,j] = 1) & \text{ else if } e[i,j] = 1\\ 
				P( \latentedge_{ikj} | e[i,j] = 0) & \text{ otherwise} \\
				\end{cases}$
			}
			pNew[i,k] = $\frac{\sum_{j = 1}^N \latentedge_{ikj}}{N-1}$\;
		} 
	}
	maxDiff = max($|$ pNew - p $|$)\;
	p = pNew
 }
 \Return(p) \
\caption{Simple EM Algorithm}
\label{alg:simple}
\end{algorithm}
\subsection{Unknown Edge Status}

In many applications, the edge status between two nodes may be unknown,
and one may wish to predict missing edges from the data. This can also be used to assess
model fit by evaluating edge predictions on masked edges. With our EM algorithm, this 
can be easily accommodated.

If we define 

\begin{equation}\label{eq:mask}
M(i,j) = 
\begin{cases}
0 & \text{if edge status for node pair $i,j$ is known} \\
1 & \text{if edge status for node pair $i,j$ is missing}, \\
\end{cases}
\end{equation} 
then we can redefine the M-step update as 
\begin{equation}\label{eq:maskUpdate}
\hat p_{ik} = \frac{ \sum_{j \neq i}^{N_n} \latentedge_{ikj} (1 - M(i,j) )}{ \sum_{j \neq i} (1 - M(i,j)) }.	
\end{equation}

\subsection{Efficient Updates}
 
While the algorithm described in Algorithm \ref{alg:simple} is straightfoward, 
many of the computations in this algorithm are redundant and the order of complexity 
of this algorithm can be reduced by caching various statistics. 

Let $E_i$ to be the set of nodes that are known share an edge with 
node $n_i$ and $E_i^c $ to be the set of nodes that are known to lack an edge with node $i$. We explicitly 
store $E_1,...,E_{N_n}$ in a list but do not explicitly store $E_i^c$. Note that
$n_i$ is neither in $E_i$ nor $E_i^c$, and if the edge status between node $n_i$ and $n_j$ is unknown, 
$n_j$ appears in neither set. 

We first note that the EM steps can be combined in the form 
\begin{equation} \label{eq:combineEM}
 p^{new}_{ik} =  
 	\frac{ \displaystyle  \sum_{j \in E_i^c} P( \latentedge_{ikj} | e_{ij} = 0) + \sum_{j \in E_i } P( \latentedge_{ikj} | e_{ij} = 1) } {|E_i| + |E^c_i|}  .
\end{equation}
Defining $M_i$ to be the set of nodes such that edge status with node $n_i$ is unknown, 
the first term of the numerator can then be rearranged as 
\begin{equation} \label{eq:noEdgeCont}
\begin{split}
\displaystyle \sum_{j \in E_i^c} P( \latentedge_{ikj} | e_{ij} = 0)  & = \sum_{j \in E_i^c} p_{ik} - p_{ik} p_{jk} \\
 & = N_n p_{ik} (1 - \bar{p}_{.k} ) -  p_{ik} \left( (1 - p_{ik})  + \sum_{j \in E_i \cup M_i} (1 - p_{jk})   \right),
 \end{split}
\end{equation} 
where $\bar p_{.k}$ represents the column mean of the matrix $p$.

Assuming $| E_i ^c | > | E_i \cup M_i |$, this reduces the computation required to compute the first term from $O( | E_i ^ c | )$ to $O( | E_i \cup M_i |)$
as long as $\bar p_{.k}$ is cached. 
%Because the ECM algorithm only updates one entry of $p$ at a time, each update 
%only requires $O(1)$ operations to update the cached $\bar p_{.k}$ at the end of each update. 

Next, if we define 
\begin{equation} \label{eq:edgeProb2}
\pi_{ij} \equiv P(e_{ij} = 1) = 1 - \prod_{k = 1}^K( 1 - p_{ik}p_{jk} )
\end{equation}
we can write 
\begin{equation} \label{eq:edgeCont}
\begin{split}
\sum_{j \in E_i } P( \latentedge_{ikj} | e_{ij} = 1)  & 
             = \sum_{j \in E_i } \left( \frac{ p_{ik} p_{jk} + p_{ik} (1 - p_{jk}) \left( 1 - \prod_{k' \neq k} ( 1 - p_{ik'} p_{jk'} ) \right) } {\pi_{ij} }  \right) \\
	&  = \sum_{j \in E_i } \left( \frac{ p_{ik} p_{jk} + p_{ik} (1 - p_{jk}) \left( 1 - \frac{ 1 -\pi_{ij} }{ 1 - p_{ik} p_{jk} } \right) }{ \pi_{ij} } \right). \\
\end{split}
\end{equation}

Using cached values of $\pi_{ij}$ reduces the computations required for the second term of
Equation \ref{eq:combineEM} from $O(K | E_i |)$ to $O( | E_i | )$. Precomputing all $\pi_{ij}$ 
within the edge list $E$ requires $O(KN_e)$ time. 
% If a single entry of $p$ is updated, we can update $\pi_{ij}$ in $O(1)$ time by computing 
% \begin{equation}
% \pi^{new}_{ij} = 1 - \frac{ (1 - \pi_{ij} )( 1 - p^{new}_{ik} p_{jk} ) } { 1 - p^{old}_{ik} p_{jk} }. 
% \end{equation}
%One technical note is that because we are considering an undirected graph, $\pi_{ij} = \pi_{ji}$ by definition. 
%This implies that if we update $p_{ik}$, we must update both cache edge probabilities $\pi_{ij}$ and $\pi_{ji}$
%unless they are explicitly saved and accessed as a single value. If $\pi_{ij}$ is stored as a sparse matrix, 
%this can be somewhat challenging to do in $O(1)$ time. 
% We addressed this issue by storing the 
% value of $\pi_{ij}$ as a probability list $P$, where $P[i][j*]$ is the edge probability between node $i$ and node $i$'s 
% $j^{th}$ edge. We also created a mapping in advance that links $P[i][j*]$ to its corresponding transpose value, 
% so that $\pi_{ij}$ and $\pi_{ji}$ can be updated in $O(1)$ time. 

Similar to \cite{bkn}, it should be noted that if $p_{ik} = 0$, then the EM algorithm will leave $p_{ik}$ unchanged. 
This can be exploited for additional speedup by skipping the update for $p_{ik}$ if $p_{ik} < \epsilon_p$
for a preset tolerance level $\epsilon_p$. 

Finally, it should be noted that while updates of a given row of the matrix $p$ will share the same $\pi_{ij}$, 
the computations across rows are independent. Thus, it is simple to parallelize the computation by
splitting up the updates per thread by row of $p$. 

Pseudocode for our full EM algorithm is shown in Algorithm \ref{alg:cache}. 
The initial computational complexity of each step of this algorithm is $O(K(N_n + N_e + N_m ) )$, where $N_m$ the total number of unknown edges,
but the complexity of later steps of the algorithm can be reduced by skipping updates where $p_{ik} < \epsilon_p$. 
\begin{algorithm}
\SetAlgoLined
\KwResult{Fixed point estimate of $N \times K$ matrix $p$}
 Edge list $E$ s.t. $E[i][j] \equiv j^{th}$ index of node sharing $j^{th}$ edge with node $i$\;
 Missing list $M$ s.t. $M[i][j]  \equiv j^{th}$ index of node sharing $j^{th}$ with unknown edge status with node $n_i$\;
 $pNew$ = RandomUniform(min = 0, max = 1, nrow = N, ncol = K)\;
 maxIters = 10,000; iter = 0\;
 tol = $10^{-4}$; pTol = $10^{-10}$; maxDiff = tol + 1\;
 p = pNew\;
 \While{ iter $<$ maxIters \& tol $>$ maxDiff}{
 	pBar = ColumnMeans(p)\;
	iter++\;
 	\For{ i in 1:N (in parallel)}{
		nEdges = length(E[i])\;
		edgeProbs = vector(nEdges)\;
		\For{ ii in 1:nEdges }{
			j = E[i][ii]\;
			edgeProbs[ii] = computeEdgeProb(i, j, p)\;
		}
		\For{ k in 1:K }{
			pik = p[i,k]\;
			\If{pik $<$ pTol}{
				skip\;
			}
			edgeSum = 0.0\;
			noEdgeSum = N * pik * (1 - pBar[k]) - pik *(1 - pik)\;
			nMissing = length(M[i])\;
			\For{ii in 1:nMissing}{
				pjk = p[M[i][ii], k]\;
				noEdgeSum -= pik * (1 - pjk)\;
			}
			\For{ii in 1:nEdges}{
				pjk = p[E[i][ii],k]\;
				noEdgeSum -= pik * (1 - pjk)\;
				ep = edgeProbs[ii]\;
				edgeSum += pik * (pjk+(1-pjk) * (1-$\frac{\text{1- ep}}{ \text{1-pik * pjk} }$ ) ) / ep\;
			}
			pNew[i,k] = (edgeSum + noEdgeSum) / (N - nMissing - 1)\;
		}
		maxDiff = max($|$ pNew - p $|$)\;
		p = pNew;	
	}	
}
\Return{p}\;
\caption{Efficient Algorithm}
\label{alg:cache}
\end{algorithm}

\subsection{Missing Edges with BKN Model}
 
To quantitatively compare the LCN model to the BKN model, we compare AUCs on missing link predictions for which 
the edge status is masked from the model. 
In \cite{bkn}, the EM algorithm presented assumes all edge statuses are known, and unlike the LCN model, 
we were not able to derive a simple closed form solution to the M-step when the edges statuses are unknown. 
However, because the unknown edge statuses appear in the complete data likelihood function in a linear manner, 
we can simply add an imputation step to their algorithm that imputes the expected edge count of the unknown edges. 
Notably, this does not increase the computational complexity of the algorithm when compared with our LCN algorithm, 
as imputing the missing edges is $O(KN_m)$, so the standard update is $O(K(N_n + N_e + N_m))$, similar to our algorithm. 
 
\section{Applications}
\label{sec:applications}

\subsection{Evaluating Performance}

We compared the LCN and BKN models on a variety of synthetic and real datasets. 
We assessed performance in both a quantitative and qualitative manner. 

For quantitative assessment, we randomly masked the edge status of 500 node pairs that shared an edge 
and 500 node pairs that lacked an edge during fitting and then predicted the edge status of node pairs in both sets. 
We compared the area under the curve of the receiver operating characteristic curve (AUC), since this measure relates to the predictive power of the model without being overly sensitive to 
overfitting as metrics such as out-of-sample log-likelihood are. Indeed, although it was very rare, both models occasionally 
estimated that a masked edge had zero probability, leading to an out-of-sample log-likelihood of $-\infty$, as one should expect for such a 
high dimensional problem fit with maximum likelihood. 

For qualitative assessment, we visually examined the fit to see if the LCN model captured relations we would 
expect from metadata about the nodes. This was done by creating a heatmap of $\hat p$. 
To assist in visualizing whether categorical metadata was associated with given channels, we first reordered the rows 
of the parameter matrix by the node's metadata and then reordered the columns of the matrix 
by the ratio of between-group and within-group variance of the parameter values. This somewhat crudely sorts the channels 
such that those near the top of the heatmap will be mostly strongly differentially expressed by category, while those near the bottom 
will be very weakly differentiated by category.

\subsection{Stochastic Block Model}

To compare the LCN and BKN models on synthetic data, 
we simulated stochastic block models with $p_{in} = 0.5$, $p_{out} = 0.02$, 
8 blocks with 32 nodes per block. We fit both models with $1, 2, 4, \ldots, 64$ number of channels.
Note that the stochastic block model can be perfectly reproduced with 8 channels. 
For each set of channels used, 25 stochastic block models were simulated and the LCN and BKN models 
were fit with the edge status of 500 edges and 500 non-edges masked from the algorithm. 
Out-of-sample errors were calculated on these withheld sets and in-sample errors were calculated on 
500 randomly chosen edges and non-edges that were used to fit the model. Mean squared errors were computed 
on the true edge probabilities rather than the actual edge statuses themselves. 
Each setup was repeated 10 times, and the mean and standard errors of the AUC were recorded. 

\begin{figure}
\centering
\includegraphics[width = 7.5cm]{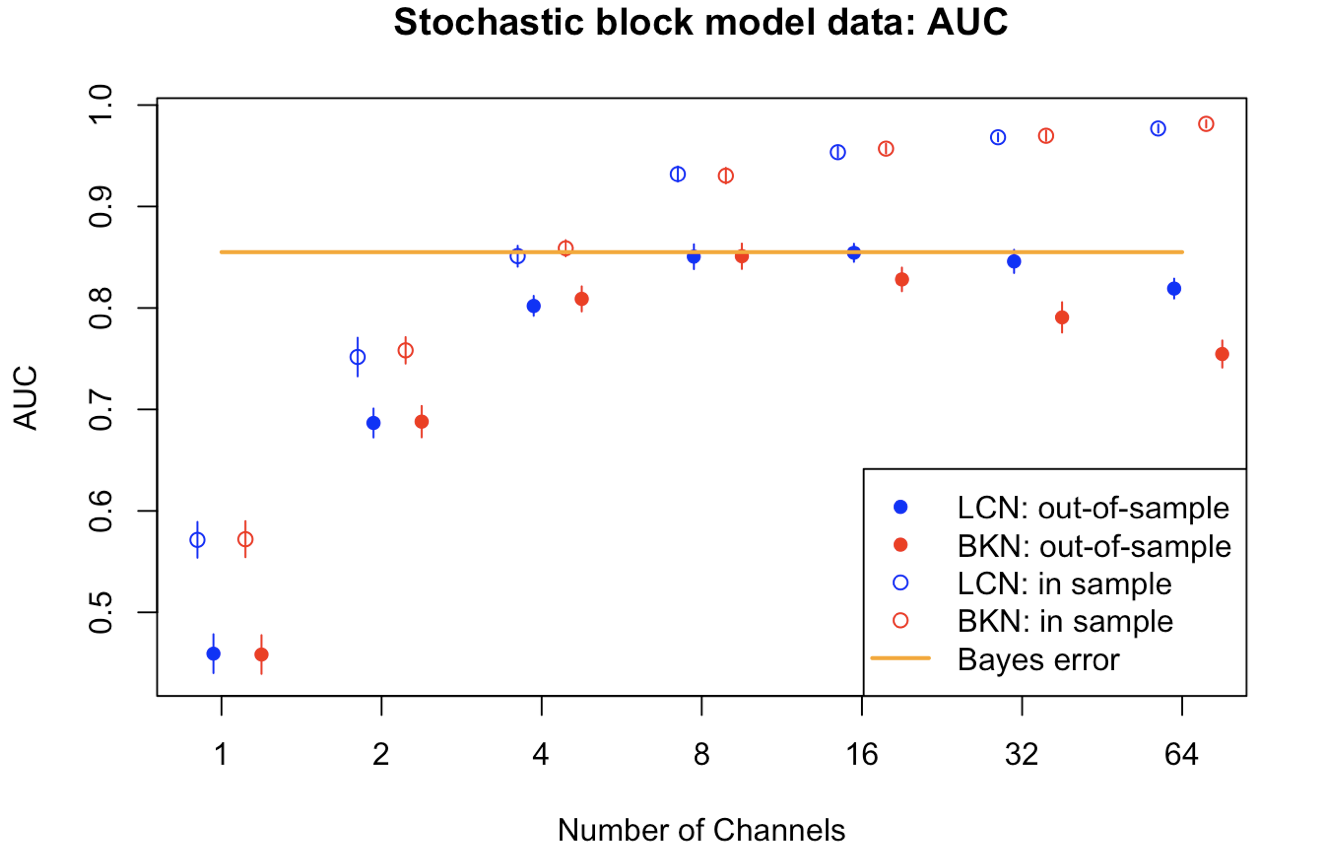}
\caption{AUC for synthetic stochastic block model.}
\label{fig:sbmAUC}
\end{figure}

%\begin{figure}
%\centering
%\includegraphics[width = 7.5cm]{sbmMSE.png}
%\caption{MSE for synthetic stochastic block model.}
%\label{fig:sbmMSE}
%\end{figure}
%
%\begin{figure}
%\centering
%\includegraphics[width = 7.5cm]{sbmRMSE.png}
%\caption{Rescaled MSE for synthetic stochastic block model. MSE is rescaled by variance associated with parameter.}
%\label{fig:sbmRMSE}
%\end{figure}

The summarized results from the simulations can be seen on Figure~\ref{fig:sbmAUC}. %, \ref{fig:sbmMSE} and \ref{fig:sbmRMSE}. 
We see that both models appear to achieve near the Bayes Error when the number of channels is equal to 
8, the minimum necessary to fully parameterize the stochastic block model. We note that the out-of-sample performance of the
LCN model appears to degrade significantly slower than the BKN model as unnecessary channels are added. 
Both models greatly overstate the AUC using the in-sample data. 
The heavy overfitting of AUCs on the SBMs is partially due to the SBMs having only two edge probabilities in the model ($p_{in}$ and $p_{out}$). 
This is described in more detail in the Appendix. 

%We hypothesis that \emph{some} of this overfitting is a peculiarity of the SBM model, which we explain in more detail in the appendix. 

An oddity that deserves some explanation is that when a single channel is used, the AUC drops below 0.5 for both models. 
Our explanation for this behavior is that with only one channel, each node has a single parameter dictating the frequency with which 
it makes connections with all other nodes. When we mask a single connection, we induce a very slight downward bias in the node pair's
parameters. Likewise, when we mask the lack of an edge, we induce a slight upward bias in the node pair's parameters. 
Since the probability of any node connecting with any other randomly sample is constant across nodes in our simple SBM example, 
this bias induced by masking makes the predictor's performance worse than that with random guessing when the number of channels is one.

\subsection{Facebook100}

For an application to real data, we considered a sample of the the Facebook 100 datasets \cite{fb100}.
These graphs contain historic Facebook data from 100 different universities. 
We picked the first ten graphs on the list for comparing optimal out-of-sample AUC and focussed on the top four for visualizations.
Basic summaries of each graph are shown in Table \ref{tab:FBsize}. Several metadata variables are available 
on each node. For the qualitative analysis, we consider enrollment year.  

% latex table generated in R 3.6.0 by xtable 1.8-4 package
% Wed Aug 21 09:38:29 2019
\begin{table}[ht]
\centering
\begin{tabular}{rlrrrr}
  \hline
 & University & Nodes & Edges & Max Degree & Median Degree \\ 
 \hline
 1 & Harvard & 15,126 & 1,649,234 & 1,183 & 77 \\ 
  2 & Columbia & 11,770 & 888,666 & 3,375 & 56 \\ 
  3 & Stanford & 11,621 & 1,136,660 & 1,172 & 72 \\ 
  4 & Yale & 8,578 & 810,900 & 2,517 & 75 \\ 
  5 & Cornell & 18,660 & 1,581,554 & 3,156 & 64 \\ 
  6 & Dartmouth & 7,694 & 608,152 & 948 & 62 \\ 
  7 & UPenn & 14,916 & 1,373,002 & 1,602 & 73 \\ 
  8 & MIT & 6,440 & 502,504 & 708 & 56 \\ 
  9 & NYU & 21,679 & 1,431,430 & 2,315 & 50 \\ 
  10 & BU & 19,700 & 1,275,056 & 1,819 & 51 \\ 
   \hline   
\end{tabular}
\caption{Size of Facebook 100 Graphs} 
\label{tab:FBsize}
\end{table}

% latex table generated in R 3.6.0 by xtable 1.8-4 package
% Sat Aug 31 14:38:08 2019
\begin{table}[ht]
\centering
\begin{tabular}{c c  c  c  c  c }
  \hline
  & & \multicolumn{2}{ c }{LCN} & \multicolumn{2}{ c }{BKN}\\
 &University & AUC & Channels & AUC & Channels \\ 
  \hline
1 & Harvard & 0.9653 & 256 & 0.9534 & 64 \\ 
 2&  Columbia & 0.9593 & 256 & 0.9466 & 64 \\ 
  3&Stanford & 0.9610 & 256 & 0.9475 & 64 \\ 
  4& Yale & 0.9597 & 256 & 0.9473 & 128 \\ 
 5& Cornell & 0.9562 & 256 & 0.9441 & 128 \\ 
 6&  Dartmouth & 0.9561 & 256 & 0.9436 & 64 \\ 
  7& UPenn & 0.9602 & 256 & 0.9461 & 64 \\ 
  8& MIT & 0.9575 & 256 & 0.9435 & 64 \\ 
  9& NYU & 0.9496 & 256 & 0.9372 & 64 \\ 
  10& BU & 0.9452 & 256 & 0.9297 & 64 \\ 
   \hline
\end{tabular}
\caption{Optimal AUC results.} 
\label{tab:optauc}
\end{table}

\begin{figure}
\centering
\includegraphics[width = \plotSize]{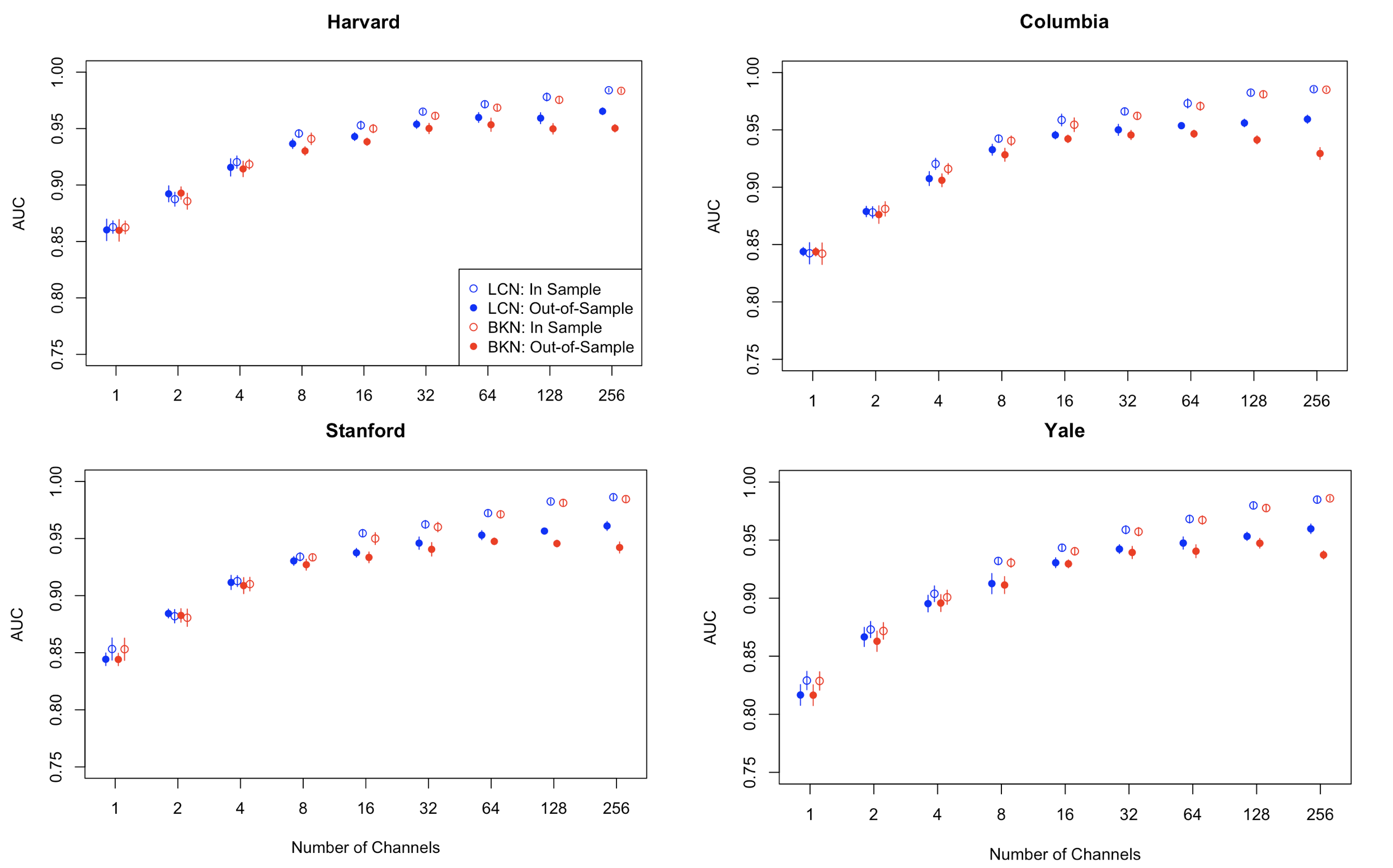}
\caption{AUC for top four datasets from Facebook 100}
\label{fig:fbauc}
\end{figure}

We fit both models with $1, 2, 4, \ldots , 256$ channels and masked 500 edges and non-edges to estimate an out-of-sample AUC. 
Each setup was repeated 10 times, and the mean and standard error of the AUC, both in-sample and out-of-sample, was recorded.
The optimal out-of-sample AUC and number of channels for the different networks can be found in Table \ref{tab:optauc}.
The AUC's by number of channels for the first four universities
are shown on Figure~\ref{fig:fbauc}. 
We make several observations:

\begin{itemize}

\item Both models were able to achieve out-of-sample AUC's over 0.929 on all ten networks. 
The best performance was usually seen with a large number of channels.

\item The performance of the BKN and LCN models was nearly identical on the Facebook networks when the number of channels 
was small to moderate ($\leq 32$ channels).

\item In larger models, ($ \geq 64$ channels), the BKN models consistently showed more overfitting when compared with 
the LCN model. This consistently led to the highest out-of-sample AUC for LCN by a moderate amount. 
Note that the optimal number of channels for the LCN model was consistently more than for the optimal number for the BKN model. 
In fact, in all ten networks, the optimal number of channels for the LCN model was the maximum number fit (256), 
although the improvement over 128 channels was minimal. 

\item Both models had moderate predictive power (out-of-sample AUC $>$ 0.8) with a single channel, which is 
the opposite of what we saw with our synthetic SBM. 
This can be explained by the skew of the degree distribution of the real networks, 
compared with a constant expected degree in the SBM graphs.

\end{itemize}

Next, we fit each network with the full edge list and plotted heat maps of the estimated $p$ matrix. 
For visualization purposes, we used 32 rather 256 channels which can be seen on Figure~\ref{fig:fbpmat}. 
Plots with 256 channels can be found in the Appendix on Figure~\ref{fig:c256mat}.
Table~\ref{tab:chanusage} shows the average usage of channels by year of enrollment, where we count a node as using a channel if the $\hat p$ entry is greater than 0.01. 
Across all four networks, we see some common trends:

\begin{itemize}

\item Many of the channels were strongly tied to year of enrollment. 
Some channels were essentially exclusively used by a cohort. 
Others were strongly used by multiple cohorts, 
with the upper cohorts using more frequently than the lower cohorts. 

\item With rare exceptions, freshmen (2009) used the channels associated 
with the freshmen class \emph{only}. 
This differed with upperclassmen, 
who used both channels attached strongly to their cohort and channels strongly used by many cohorts. 
Sophomores tended to use multi-cohort channels less than juniors and seniors did, but more than freshmen did. 

\item Freshmen, and to a lesser extent sophomores, tended to make edges through fewer channels than juniors and seniors. 
%Table~\ref{tab:chanusage} shows the average usage of channels by year of enrollment, where we count a node as using a channel if the $\hat p$ entry is greater than 0.01.  

\item Overall, the $\hat p$ matrix was fairly sparse. 
Across the four networks, the proportion of exact zeros in $\hat p$ ranged from 0.839 and 0.860.
When the model was fit with more channels, $\hat p$ was more sparse, as expected. 

\end{itemize}

% latex table generated in R 3.6.0 by xtable 1.8-4 package
% Sun Sep  1 15:36:15 2019
\begin{table}[ht]
\centering
\begin{tabular}{rrrrrr}
  \hline
 & 2006 & 2007 & 2008 & 2009 & All Classes \\ 
  \hline
Harvard & 5.80 & 5.18 & 4.25 & 2.04 & 3.88 \\ 
  Columbia & 4.55 & 4.80 & 4.20 & 2.40 & 3.67 \\ 
  Stanford & 5.76 & 5.91 & 4.92 & 2.04 & 4.09 \\ 
  Yale & 5.77 & 5.67 & 5.38 & 2.73 & 4.38 \\ 
   \hline
\end{tabular}
\caption{Average number of channels used per node by cohort.} 
\label{tab:chanusage}
\end{table}

\begin{figure}
\centering
\includegraphics[width = \plotSize]{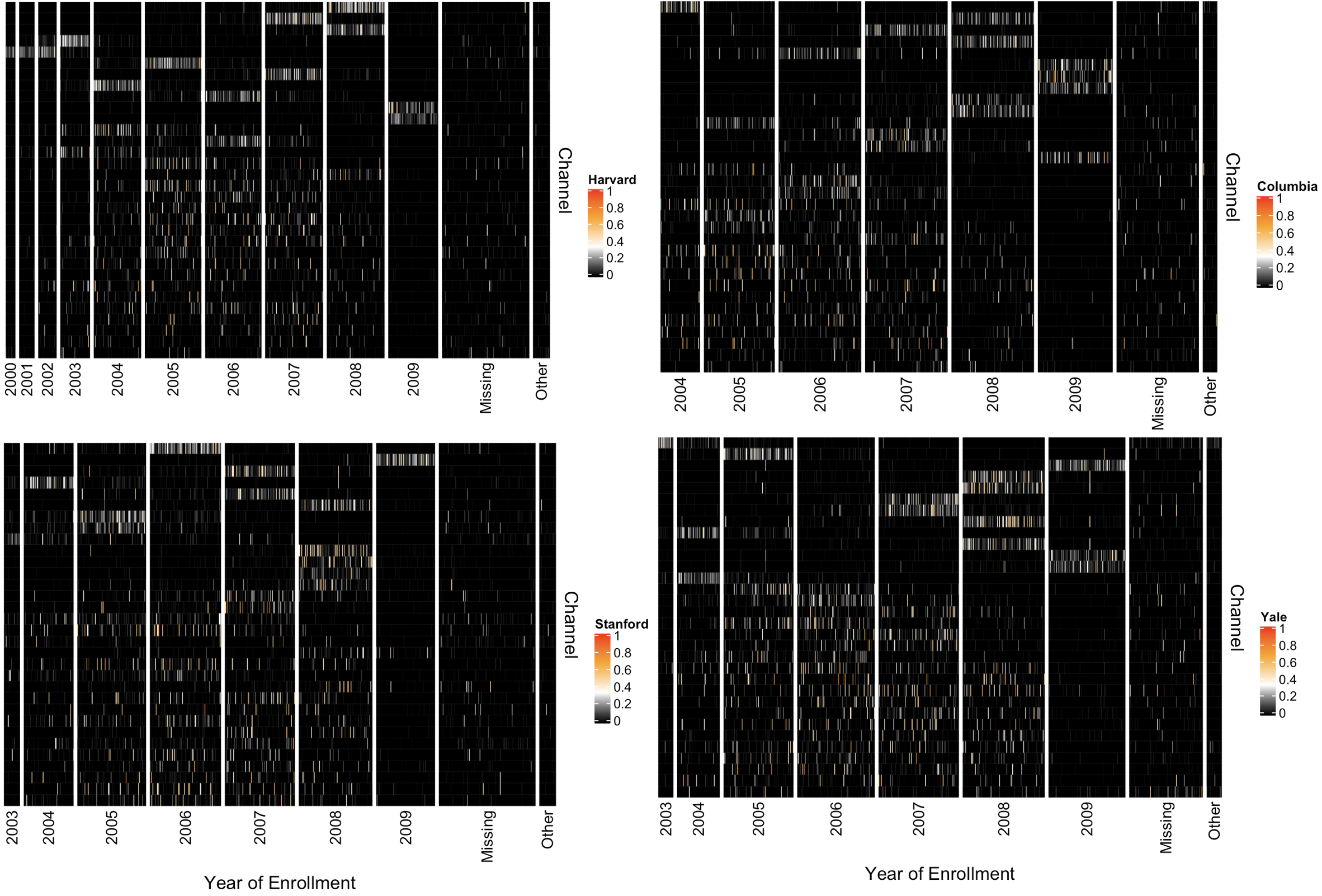}
\caption{Heatmaps for estimated $p$ matrix for each Facebook 100 dataset. Thirty-two channels were used for visualization purposes.}
\label{fig:fbpmat}
\end{figure}

We note that these results seem consistent with what one might expect from a university network; 
freshmen are very likely to make friends in their lower division courses, whereas by the time students 
become upperclassmen, many friendships and groups will haved formed that are less strictly determined by cohort. 

\subsection{Synthetic Latent Channel Graphs}

Interestingly, our real data examples demonstrated significantly higher out-of-sample AUC than the 
Bayes Error of the SBM example we evaluated earlier. 
Initially, this was slightly perplexing, as the parameters that we specified for the SBM 
($ p_{in} = 0.5$, $p_{out} = 0.02$) implied quite strong individual communities. 
In addition, the synthetic data appeared to show more issues with overfitting than the real data. 

While it was encouraging to learn that our model worked better on real data than synthetic data, 
it was not immediately clear \emph{why} both the LCN and BKN models performed better on the Facebook100 data 
than on stochastic block model data with strong communities. 
After reviewing the fitted models, we pinpointed two factors that seemed to contribute to better performance of the LCN and BKN models:

\begin{itemize}

\item Sparse $p$ matrix. 

\item Small number of large degree nodes.

\end{itemize}

\begin{figure}
\centering
\includegraphics[width = 9cm]{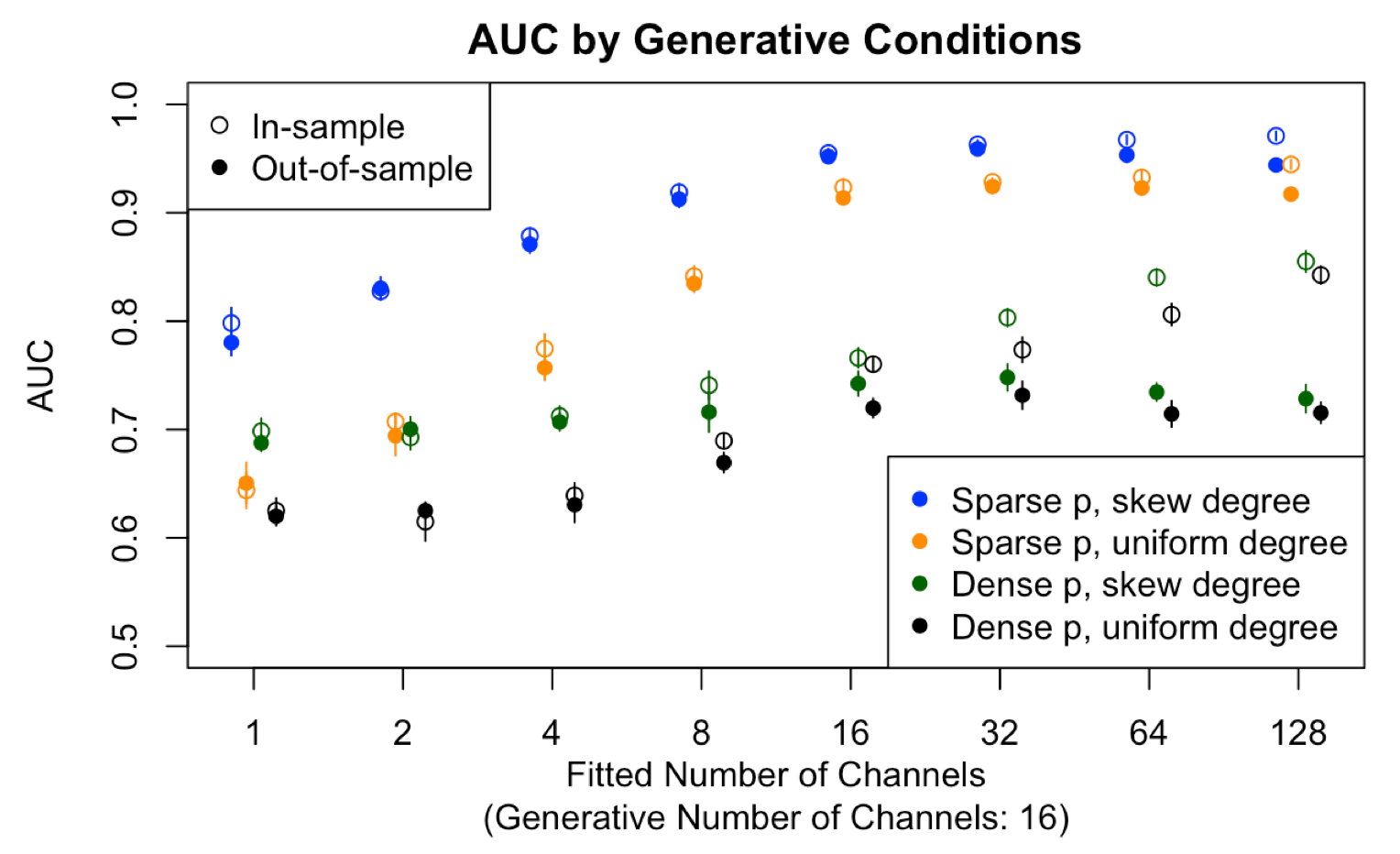}
\caption{Simulated in-sample and out-of-sample AUC based on different generative Latent Channel Network $p$ structures.}
\label{fg:genauc}
\end{figure}

To demonstrate this, we simulated data according an LCN generative model under
different conditions; sparse $p$ vs dense $p$ and skewed degrees vs uniform degrees.
The conditions dictated how the parameters of the $p$ matrix were drawn, which was then used to a draw a graph according to the LCN model. 
Associated with each node were two types of channel usages: \emph{main channels} and \emph{background channels}.
To simulate sparse $p$'s, all the background channels were set to 0. With dense $p$'s, the background channels follow a Beta($a = 1, b = 20$).
To simulate skewed degrees, the number of main channels is simulated as 1 + Beta-binomial($a = 1, b = 10, n = 15$). 
With the relatively uniform degrees, all nodes have 3 main channels. The main channels followed a Uniform(0,1) distribution. 
Each simulated graph had 1,000 nodes and 16 channels. After the data was simulated, it was fit with an LCN model with $1, 2, 4, \ldots, 128$ channels. 
Each scenario was repeated 10 times and the mean and standard error of the in-sample and out-of-sample AUC was recorded.
The results were plotted on figure \ref{fg:genauc}. We make several notes about this results.

\begin{itemize}

	\item The out-of-sample AUC for sparse, skewed degree scenario was quite similar to the results in the Facebook100 datasets. 
	
	\item With a dense $p$ matrix, the optimal out-of-sample AUC was poor, never doing much better than 0.7. 
	
	\item With a skew degree, fits with only a single channel still had some predictive power, despite being a heavily under-parameterized model. 
	We hypothesize this is the single channel being used to identify the high degree nodes. 
	
	\item With sparse $p$ and skewed degrees, the model was fairly resistant to overfitting. 
	The model was fully parameterized with 16 channels, 
	yet  little loss in out-of-sample AUC was seen when 32 and 64 channels were used to fit the data. 
	Only at 128 channels was a moderate drop in AUC observed. 

\end{itemize}

These observations suggest some serendipity on our part. 
Data may be exactly generated by a Latent Channel Network model and fit with the correct dimensional space, 
yet the fitted model can still lead to poor predictive behavior for dense $p$ matrices. 
The fact that the Facebook100 datasets appear well approximated by a sparse $p$ matrix with a highly skewed degree distribution 
implies they are the type of graphs for which the unknown edge status can be predicted well. 

\section{Discussion}
\label{sec:discuss}

We have presented the latent channel network, 
a model that allows nodes to share an edge if they connect through 
\emph{at least} one unobserved channel, which we believe captures
an important aspect of social networks. We implemented an EM algorithm 
that scales linearly in the number of edges in the graph and number of
channels in the model. 
We applied this model to the top ten Facebook100 networks and consistently found a moderate improvement 
in predicting unknown edge status in comparison with the BKN model.
We found the channels uncovered by this model tended to correspond 
with meaningful features of the data and gave insight into the structure of the graph. 

An R \cite{R} implementation of the algorithm has been made available in a GitHub repo \cite{myRepo}. 
The code uses Rcpp \cite{Rcpp} and RcppParallel \cite{RcppParallel} for efficient computations. 

There are several ways that this work can be further expanded. 
In regards to efficient computation, although each iteration of the EM algorithm is relatively 
computationally cheap, typically, several thousand iterations are
required. There are several ways in which the EM-algorithm can be accelerated.
One generic method is the SquareEM algorithm \cite{squarem} for 
reducing the iterations required until convergence. Unfortunately, this algorithm 
requires computating the observed log-likelihood. Given that each iteration of our 
cached EM-algorithm requires $O(KN_e)$ observations and 
computation of the observed log-likelihood requires $O(K N_n^2)$, 
this approach is likely to only reduce wall-time computations (rather than iterations) 
if we have a dense graph with $N_e \approx N_n^2$. 
A more promising approach is using efficient data augmentation algorithms \cite{dataAug} \cite{yuSqueeze}, 
in which the missing data  is \emph{less} informative but still provides a closed form 
solution, reducing the number of iterations required without 
significantly increasing the computational cost per iteration. 
This approach requires clever data augmentation schemes and while several have 
been proposed for mixture models, it is not obvious how 
our problem may be viewed as that of a mixture model, so novel data imputation methods 
would be required to specialize to our problem. 

%To improve the model, we may consider a 
%hybrid approach with Euclidean embedding. 
%To motivate this, consider a simplified example of 
%communication between researchers at universities. 
%Suppose the two largest driving factors
%for communication are that they  do research in the same field of 
%study (but potentially different universities) and that they  belong
%to the same university (but at potentially different field of studies). 
%This could be represented as a latent channel model, but we would need one channel 
%for each university and each field of study. Alternatively, suppose 
%we considered multiple latent Euclidean spaces, 
%and our model allowed for an edge if there was a connection through 
%\emph{at least} one of these spaces. We would have 
%one space representing university and one space representing field of study. 
%Thus, researchers at the same university could be close in the two-dimensional embedded space representing
%university and likewise for field of study. 
%Note that this would require only four parameters per subject for the embedding 
%(plus two more if an intercept in each space was included),
%regardless of the number of universities or fields of research.
%We believe this should greatly reduce the size of the 
%parameter space required to describe complex networks, 
%although estimating parameters for this 
%model is likely to be quite challenging. 

\section{Acknowledgements}

We would like to thank Goran Konjevod for informative discussions 
on graph algorithms. 
All heatmaps in this paper were generated by \cite{complexHeatmaps}. 

\section{Appendix}
\subsection{Bayes Error for Stochastic Block Model}

One definition of the AUC measure is 
\begin{equation}
P(\hat y > \hat y' | y = 1, y' = 0).
\end{equation} 
In terms of edge prediction, we can write this as 
\begin{equation}
P(\hat e_{ij} > \hat e_{i'j'} | e_{ij} = 1, e_{i'j'} = 0).
\end{equation} 
To compute the Bayes error for Stochastic Block Models, we clarify how to calculate this metric with ties in the predictor as
\begin{equation}
P(\hat e_{ij} > \hat e_{i'j'} | e_{ij} = 1, e_{i'j'} = 0) + \frac{1}{2} P(\hat e_{ij} = \hat e_{i'j'} | e_{ij} = 1, e_{i'j'} = 0)
\end{equation}
Note that this definition is equivalent to the previous, where ties in the the estimator are broken up at random. 
In a stochastic block model, the Bayes estimator for an edge is an indicator function of whether 
the node pair belongs to the same block. We define this as $B(i,j)$, where $B(i,j) = 1$ if nodes $i$ 
and $j$ are from the same block and $B(i,j) = 0$ if they are from different blocks.
We consider the case in which 
$p_{in}$, $p_{out}$ and the block size is the same across all blocks. If we define
$n_b$ as the number blocks and $n_s$ as the size of the blocks and
$\pi_B \equiv P(B(i,j) = 1) = \frac{n_s -  1}{n_b n_s - 1} $,
we can write the AUC as 
\begin{equation}
\frac{  P(B(i,j) > B(i',j') , e_{ij} = 1, e_{i'j'} = 0) + \frac{1}{2} P(B(i,j) = B(i',j') , e_{ij} = 1, e_{i'j'} = 0)  }{ P( e_{ij} = 1, e_{ij} = 0) }.
\end{equation}
To evaluate and simplify this equation, we define $q_{in} = 1 - p_{in}$, $q_{out} = 1 - p_{out}$ and $q_{B} = 1 - \pi_{B}$.
Then the Bayes Error for the AUC can be written in closed form as
\begin{equation}
\frac{  
p_{in} q_{out}  \pi_B q_B + 
\frac{1}{2} (p_{in} q_{in} \pi_B^2 + 
p_{out} q_{out} q_B^2 )
}{ 
p_{in} q_{out}  \pi_B q_B + 
p_{in} q_{in} \pi_B^2 + 
p_{out} q_{out} q_B^2 +
p_{out} q_{in}  q_B \pi_B
}.
\end{equation}

\subsection{Overfitting AUC with SBMs}

In the applications section, it was observed that the in-sample AUC greatly estimated the predictive power of both the LCN and BKN models. 
We note that the AUC is defined as 
\begin{equation}
P(\hat e_{ij} > \hat e_{i'j'} | e_{ij} = 1, e_{i'j'} = 0) + \frac{1}{2} P(\hat e_{ij} = \hat e_{i'j'} | e_{ij} = 1, e_{i'j'} = 0).
\end{equation}
If we know the predictor $\hat e_{ij}$ only up to a very small amount of continuous noise, all the ties in the data would be broken, 
and the estimated AUC would be 
\begin{equation}
P(\hat e_{ij} > \hat e_{i'j'} | e_{ij} = 1, e_{i'j'} = 0) + P(\hat e_{ij} = \hat e_{i'j'} | e_{ij} = 1, e_{i'j'} = 0).
\end{equation}
Because a simple SBM only has two unique edge probabilities, a very large number of ties in the true edges probabilities exist. 
Since the LCN and BKN models do not constrain any of the edge probabilities to be exactly equal, 
\emph{any} error in the estimated edge probabilities automatically upwardly bias the in-sample AUC by 
$ \frac{1}{2} P(\hat e_{ij} = \hat e_{i'j'} | e_{ij} = 1, e_{i'j'} = 0)$.
 
\subsection{Extending BKN Algorithm for Unknown Edge Status}

The algorithm presented by Ball, Karrer and Newman can be presented as an EM algorithm, 
in which the E-step consists of computing

\begin{equation}
q_{ijk} = \frac{ \theta_{ik}\theta_{jk}}{ \sum_k \theta_{ik}\theta_{jk} }
\end{equation}

and the M-step consists of computing

\begin{equation}
\theta_{ik} = \frac{ \sum_j A_{ij} q_{ijk} } { \sqrt{\sum_{ij} A_{ij} q_{ijk} } }.
\end{equation}

If the edge status of a set of node pairs is unknown, then the value of $ A_{ij} $ will be unknown
for a set of $ij$ pairs. To account this, we add a step to estimate the missing data with

\begin{equation}
\tilde A_{ij} = \sum_{k} \theta_{ik} \theta_{jk}.
\end{equation}

and use $\tilde A_{ij}$ for the unknown edges in the M-step.

\subsection{Plots of $\hat p$ with 256 Channels}

\begin{figure}
\centering
\includegraphics[width = \plotSize]{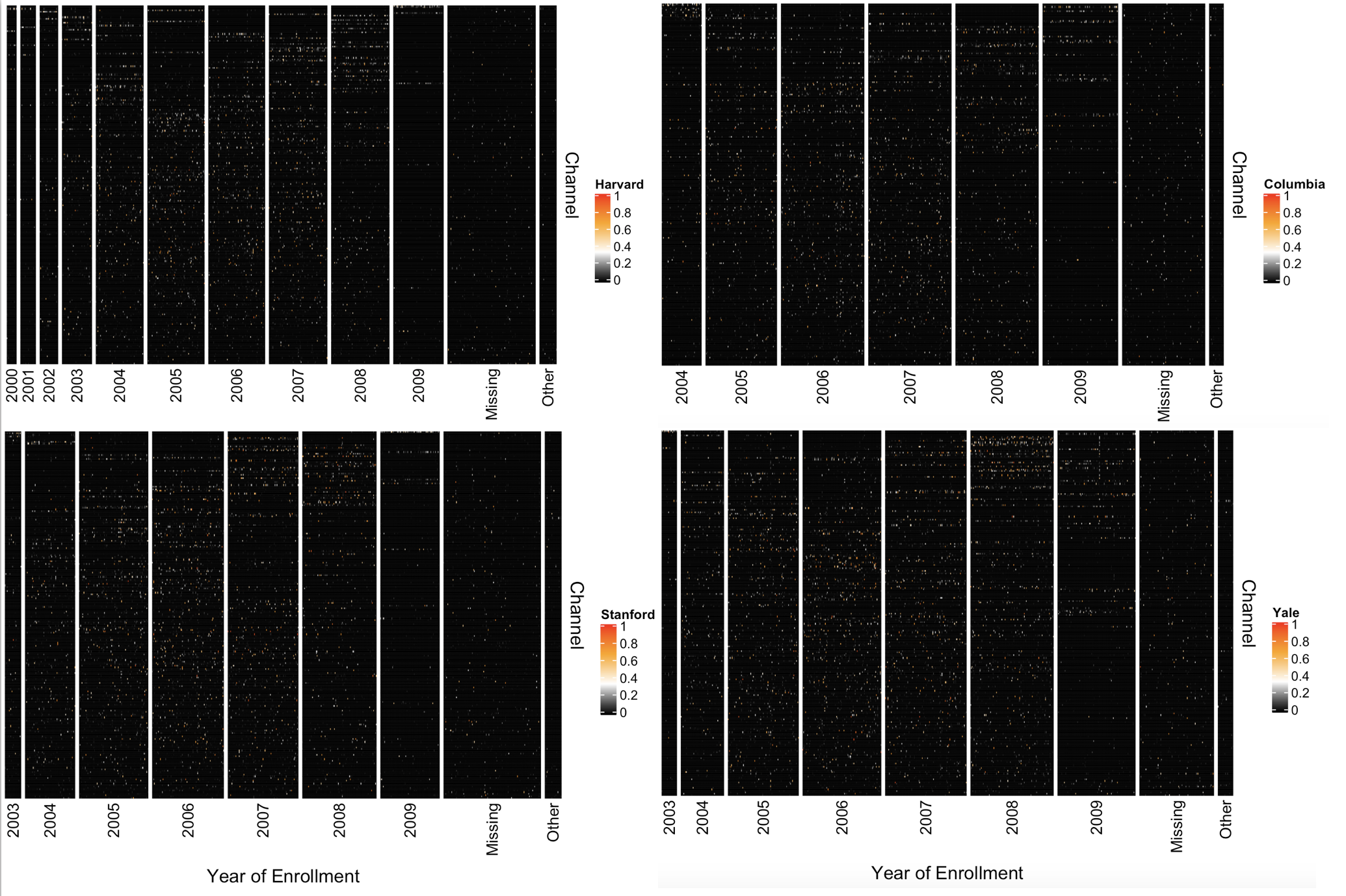}
\caption{Heatmaps for estimated $p$ matrix for each Facebook 100 dataset with 256 channels. }
\label{fig:c256mat}
\end{figure}

In Figure \ref{fig:c256mat}, we show the $\hat p$ matrix with 256 channels. 
We chose not to focus on these plots as the minute cell size makes visualization difficult. 
If one wished to visualize $\hat p$ for such a large matrix, 
we suggest focusing on a subset of interesting channels.

\bibliographystyle{siam}
\bibliography{LatentChannels.bib}

\end{document}